\documentclass[a4paper,12pt]{article}

\usepackage{amsmath}
\usepackage{amsthm}
\usepackage{amssymb}
  \newtheorem{df}{Definition}[section]
  \newtheorem{thm}[df]{Theorem}
  \newtheorem{lem}[df]{Lemma}
  \newtheorem{rmk}[df]{Remark}
  
  \newtheorem{cor}[df]{Corollary}
\makeatletter
   
   \@addtoreset{equation}{section}
  \makeatother

\title{A Generalization of the Submodel of \\
Nonlinear ${\mathbf{C}}P^1$ Models}
\author{Tatsuo SUZUKI\thanks{Department of Mathematical Sciences, 
  Waseda University, 
  Tokyo 169-8555, 
  Japan, \endgraf 
  {\it E-mail address}: suzukita{\char'100}mse.waseda.ac.jp}}
\date{}
\begin{document}
\maketitle
\begin{abstract}
We generalize the submodel of nonlinear ${\mathbf{C}}P^1$ models. The generalized models include higher order derivatives. For the systems of higher order equations, we construct a B\"acklund-like transformation of solutions and an infinite number of conserved currents by using the Bell polynomials. 
\end{abstract}
\section{Introduction}
Integrable theories in a space-time of two dimension have achieved remarkable 
development and great many integrable models exist. 
But in the dimensions greater than two, 
we do not have so many interesting models because it is difficult to extend 
the concepts of integrability to higher dimensions. 

In these circumstances, O. Alvarez, L. A. Ferreira and J. S. Guillen proposed a new approach to higher dimensional integrable theories \cite{AFG1} (see also \cite{AFG2}). Instead of higher dimensional models themselves , they considered their submodels to construct integrable models in the sense of possessing 
an infinite number of conserved currents. They applied their theories to 
nonlinear ${\mathbf{C}}P^1$ model in $(1+2)$-dimensions in \cite{AFG1}. 

Then we calculated an infinite number of conserved currents explicitly 
in the submodel of nonlinear ${\mathbf{C}}P^1$ model in $(1+2)$-dimensions 
\cite{FS1},\cite{FS2} (see also \cite{GMG}). 
Furthermore, we generalized the definition of submodels to nonlinear Grassmann sigma models and constructed an infinite number of conserved currents and 
a wide class of exact solutions \cite{FHS1},\cite{FHS2}. 
(later Ferreira and Leite generalized them to 
homogeneous-space models \cite{FL}). 
The idea of submodels was also applied in \cite{AFZ1},\cite{AFZ2}. 

Moreover, we also generalized them to another direction in \cite{FHS2}. 
In view of the fact that the ${\mathbf{C}}P^1$-submodel 
$$ \partial^{\mu}\partial_{\mu}u=0 \quad \mbox{and} \quad 
 \partial^{\mu}u\partial_{\mu}u=0,
$$
$$
 \hspace{5cm} \mbox{for} \quad
 u:M^{1+n} \longrightarrow {\mathbf{C}}
$$
is equivalent to
$$
\square_2 u=0 \quad \mbox{and} \quad \square_2 (u^2)=0, 
$$
we defined a system of 
$p$-th order ($p=2,3,\cdots $) nonlinear partial differential equations (PDE) 
by generalizing the 
${\mathbf{C}}P^1$-submodel, which have an infinite number of conserved 
currents and a wide class of exact solutions;
$$
\square_p (u^k) \equiv 
  \left(
   \frac{\partial^p}{\partial x_0^p}
   -\sum_{j=1}^n \frac{\partial^p}{\partial x_j^p}
  \right) (u^k)=0 \quad \mbox{for} \quad 1 \le k \le p .
 \label{eqn:p-sub}
$$
Hereafter, we call this system of PDE {\it{the $p$-submodel}} for short. 
In constructing conserved currents of the $p$-submodel, 
we defined differential operators 
by using the Bell polynomials. We investigated the reason why such a form of 
operators appeared. Then we found a kind of ``symbol structure" for the 
operators and could define a wider class of PDE than the $p$-submodel. 

In this paper, we define a new system of PDE including the $p$-submodel 
and construct a B\"acklund-like transformation of solutions and an infinite number of conserved currents by using the Bell polynomials. 

\section{Bell Polynomials}
Firstly, we prepare a 
mathematical tool which plays an important role in our following theory. 
\begin{df}
Let $g(x)$ be a smooth function and $z$ a complex parameter.
Put $g_r \equiv \partial_x^r g(x)$. 
We define the Bell polynomials of degree $n$ (\cite{Ri}):
\begin{equation}
 F_n(zg) = F_n(zg_1, \cdots ,zg_n) 
         \equiv   \mbox{e}^{-zg(x)} \partial_x^n \mbox{e}^{zg(x)}.
\end{equation}
\end{df}
The generating function of 
Bell polynomials is 
\begin{equation}
 \exp \left\{ z\sum_{j=1}^{\infty}\frac{g_j}{j!}t^j \right\}
  = \sum_{n=0}^{\infty}\frac{F_n(zg)}{n!}t^n. 
 \label{eqn:gen}
\end{equation}
By (\ref{eqn:gen}), we can write them explicitly as follows:
\begin{equation}
F_n(zg_1, \cdots ,zg_n) = \hspace{-5mm}
    \sum_{{\scriptstyle k_1+2k_2+ \cdots +nk_n=n}\atop
          {\scriptstyle k_1 \geq 0,\cdots ,k_n \geq 0}}
         \frac{n!}{k_1! \cdots k_n!} 
         \left(
          \frac{zg_1}{1!}
         \right)^{k_1}
         \left(
          \frac{zg_2}{2!}
         \right)^{k_2}
            \cdots
         \left(
          \frac{zg_n}{n!}
         \right)^{k_n}.
\end{equation}
For example, 
$$
F_0=1, \qquad F_1=zg_1, \qquad F_2=zg_2 + z^2 g_1^2.
$$
These polynomials are used in the differential calculations of composite 
functions. 
\begin{lem}
 We have a recursion formura for the Bell polynomials.
\begin{equation}
 F_{n+1}(zg)=
  \left\{ 
   \sum_{r=1}^n g_{r+1} \frac{\partial}{\partial g_r}+zg_1
  \right\} F_n(zg).
 \label{eqn:shift}
\end{equation}
 \label{lem:recur1}
\end{lem}
Now, we define the Bell matrix 
(we have adopted notations in \cite{Al-Fre}); 
$$B_{nj}=B_{nj}[g]=B_{nj}(g_1,\cdots,g_{n-j+1})$$ 
by the following equation 
\begin{equation}
 F_n(zg_1,\cdots,zg_n)=\sum_{j \geq 0} z^j B_{nj}(g_1,\cdots,g_{n-j+1}). 
\end{equation}
Note that
\begin{equation}
B_{n0}=\delta_{n0}, \ B_{nj}=0 \ (n < j).
\end{equation}
An important formula for constructing a B\"acklund-like transformation 
is as follows;
\begin{lem}
\begin{equation}
 B_{jk}[f(g(x))]=\sum_{n=k}^j B_{jn}[g]B_{nk}[f]
 \label{eqn:Bmat}
\end{equation}
\end{lem}

 \section{A New System of Higher Order Equations}
In this section, we generalize the equations of motion of 
the ${\mathbf{C}}P^1$-submodel 
to higher order. Hereafter, we use a 
notation of Minkowski summation as follows:
\begin{equation}
 \sum_{\mu}{}'A_{\mu} \equiv A_0-\sum_{j=1}^n A_j. 
\end{equation}

Now, given $p=2,3,\cdots $ and $i=0,1,\cdots,[(p-1)/2]$, 
where [\ ] means the Gauss's symbol, 
we define a system of higher order nonlinear PDE as follows:
\begin{df}
\begin{equation}
 \sum_{\mu}{}'\partial_{\mu}^{p-i}(u^k)
   \partial_{\mu}^i(\bar{u}^l)=0
 \label{eqn:4-4}
\end{equation}
for $k=1,\cdots, p-i$, $l=0,\cdots,i$. \\
We call this system of PDE the $(p,i)$-submodel. 
\end{df}
For example, 
$(p,i)=(2,0);$
\begin{equation}
  \sum_{\mu}{}'\partial_{\mu}^2 u=0, \quad 
  \sum_{\mu}{}'\partial_{\mu}^2 (u^2)=0.
 \label{eqn:2-0sub}
\end{equation}
We note that (\ref{eqn:2-0sub}) is equivalent to the 
${\mathbf{C}}P^1$-submodel. \\
$(p,i)=(3,0);$
\begin{equation}
  \sum_{\mu}{}'\partial_{\mu}^3 u=0, \quad 
  \sum_{\mu}{}'\partial_{\mu}^3 (u^2)=0, \quad 
  \sum_{\mu}{}'\partial_{\mu}^3 (u^3)=0,
 \label{eqn:3-0sub}
\end{equation}
$(p,i)=(3,1);$
\begin{equation}
  \sum_{\mu}{}'\partial_{\mu}^2 u \partial_{\mu} \bar{u}=0, \quad 
  \sum_{\mu}{}'\partial_{\mu}^2 (u^2) \partial_{\mu} \bar{u}=0.
 \label{eqn:3-1sub}
\end{equation}

 \section{Conserved Currents for the System of Higher Order Equations}
In this section, we construct conserved currents for the system of PDE 
(\ref{eqn:4-4}).

let $g(x)$, $\bar{g}(x)$ be smooth functions and $z$, 
$\bar{z}$ complex parameters. 
Put $\cal{P}_{\mbox{B}}$ the vector space over $\mathbf{C}$ spaned by 
the products of two Bell polynomials $F_n(zg)\bar{F}_m(\bar{z}\bar{g})$. 
(We use a notation $\bar{F}_m(\bar{z}\bar{g})$ 
instead of $F_m(\bar{z}\bar{g})$ 
for convenience.) 
We consider a linear map 
\begin{equation}
 \Phi : \cal{P}_{\mbox{B}}
   \rightarrow 
   \mathbf{C}[\xi, \bar{\xi}], 
\end{equation}
\begin{equation}
 \Phi (F_n(zg)\bar{F}_m(\bar{z}\bar{g}))=\xi^n \bar{\xi}^m. 
\end{equation}
\begin{rmk}
The map $\Phi$ is considered as the tensor product of a ``symbol map" 
$$
 F_n(zg)=\mbox{e}^{-zg(x)} \partial_x^n \mbox{e}^{zg(x)}
  \longmapsto 
 \mbox{e}^{-zg(x)} {\xi}^n \mbox{e}^{zg(x)}=\xi^n . 
$$
\end{rmk}
By this map, $\cal{P}_{\mbox{B}}$ is linear isomorphic to $\mathbf{C}[\xi, \bar{\xi}]$. 
Now, we define an operator 
\begin{equation}
 \partial \equiv 
     \sum_{r=1}^{\infty} 
     \left( 
       g_{r+1} \frac{\partial}{\partial g_r}
      + \bar{g}_{r+1} \frac{\partial}{\partial \bar{g}_r}
     \right)
     +zg_1+\bar{z}\bar{g}_1
\end{equation}
This operator is well-defined on $\cal{P}_{\mbox{B}}$ and we have
\begin{equation}
 \partial (F_n(zg)\bar{F}_m(\bar{z}\bar{g}))
 = F_{n+1}(zg)\bar{F}_m(\bar{z}\bar{g})
  +F_n(zg)\bar{F}_{m+1}(\bar{z}\bar{g}).
 \label{eqn:der}
\end{equation}
In fact, because $F_n$ is $n$-variable polynomial and 
on account of (\ref{eqn:shift}), 
\begin{eqnarray}
 && \hspace{-5mm} 
    \partial (F_n(zg)\bar{F}_m(\bar{z}\bar{g})) \nonumber\\
 &=& \left\{
      \sum_{r=1}^n
        g_{r+1} \frac{\partial}{\partial g_r}
      +\sum_{r=1}^m
        \bar{g}_{r+1} \frac{\partial}{\partial \bar{g}_r}
      +zg_1+\bar{z}\bar{g}_1
     \right\}
     F_n(zg)\bar{F}_m(\bar{z}\bar{g}) \label{eqn:welldef}\\
 &=&  \sum_{r=1}^n
        g_{r+1} \frac{\partial F_n(zg)}{\partial g_r}\bar{F}_m(\bar{z}\bar{g})
       +zg_1 F_n(zg)\bar{F}_m(\bar{z}\bar{g}) \nonumber\\
  && \quad 
      +F_n(zg) 
       \sum_{r=1}^m
         \bar{g}_{r+1} 
        \frac{\partial \bar{F}_m(\bar{z}\bar{g})}{\partial \bar{g}_r}
      +\bar{z}\bar{g}_1 F_n(zg)\bar{F}_m(\bar{z}\bar{g}) \nonumber\\
 &=&  F_{n+1}(zg)\bar{F}_m(\bar{z}\bar{g})
     +F_n(zg)\bar{F}_{m+1}(\bar{z}\bar{g}). \nonumber
\end{eqnarray}
Because of (\ref{eqn:der}), we have
\begin{equation}
 \Phi \circ \partial \circ \Phi^{-1} 
 = (\xi + \bar{\xi} ),
  \label{eqn:opes}
\end{equation}
where the right hand side of (\ref{eqn:opes}) means 
the multiplication operator. 
By using the linear isomorphism $\Phi $, 
we can identify
\begin{equation}
 F_n \bar{F}_m \quad \mbox{with} \quad \xi^n \bar{\xi}^m 
 \quad \mbox{and} \quad 
 \partial \quad \mbox{with} \quad (\xi+\bar{\xi}) \ .
  \label{eqn:iden} 
\end{equation}

Nextly, we choose an $\mu \in \{ 0,\cdots,n \}$ 
and put $x=x_{\mu}$, \\
$g(x_{\mu})=u(x_0,\cdots,x_{\mu},\cdots,x_n)$.
Then, we have $g_r=\partial_{\mu}^r u$. 
We set $F_{n,\, \mu}$ as
\begin{eqnarray}
F_{n,\, \mu} 
 & \equiv & :F_n(zg_1, \cdots ,zg_n)|_{z=\frac{\partial}{\partial u}}:\\
 &=&            :F_n(\partial_{\mu}u \frac{\partial}{\partial u},
                  \partial_{\mu}^2 u \frac{\partial}{\partial u},
                   \cdots ,
                  \partial_{\mu}^n u \frac{\partial}{\partial u}):\\
 &=& \hspace{-10mm}
    \sum_{{\scriptstyle k_1+2k_2+ \cdots +nk_n=n}\atop
          {\scriptstyle k_1 \geq 0,\cdots ,k_n \geq 0}} \hspace{-1mm}
         \frac{n!}{k_1! \cdots k_n!} 
         \left(
          \frac{\partial_{\mu}u}{1!}
         \right)^{k_1} \hspace{-2mm}
         \left(
          \frac{\partial_{\mu}^2 u}{2!}
         \right)^{k_2} \hspace{-3mm}
            \cdots
         \left(
          \frac{\partial_{\mu}^n u}{n!}
         \right)^{k_n} \hspace{-2mm}
         \left(
          \frac{\partial }{\partial u}
         \right)^{k_1+k_2+ \cdots +k_n}
         \nonumber \\
   && 
\end{eqnarray}
and $\bar{F}_{n,\, \mu}$ its complex conjugate of $F_{n,\, \mu}$, where $: \ :$  means the normal ordering.

\begin{lem}
\begin{equation}
 \partial_{\mu} :F_{n,\, \mu}\bar{F}_{m,\, \mu}:f(u,\bar{u})
 \ =\ 
   :\partial (F_n(zg)\bar{F}_m(\bar{z}\bar{g}))
     |_{z=\frac{\partial}{\partial u}}:f(u,\bar{u}).
\end{equation}
where $f=f(u,\bar{u})$ is any function in $C^{n+m+1}$-class. 
\end{lem}
\textit{proof}: \ 
If $\partial_{\mu}$ acts on a functional of the form
\begin{equation}
 h(u, \partial_{\mu} u, \cdots, \partial_{\mu}^n u;\  
   \bar{u}, \partial_{\mu} \bar{u}, \cdots, \partial_{\mu}^m \bar{u}),
\end{equation}
we can write 
\begin{equation}
 \partial_{\mu} =
     \sum_{r=1}^n
       \partial_{\mu}^{r+1}u
         \frac{\partial}{\partial (\partial_{\mu}^r u)}
    +\sum_{r=1}^m
       \partial_{\mu}^{r+1}\bar{u} 
         \frac{\partial}{\partial (\partial_{\mu}^r \bar{u})}
     +\partial_{\mu}u \frac{\partial}{\partial u}
     +\partial_{\mu}\bar{u} \frac{\partial}{\partial \bar{u}}.
\end{equation}
On the other hand, in view of (\ref{eqn:welldef}) and since
$g_r=\partial_{\mu}^r u$, $z=\frac{\partial}{\partial u}$, 
we have proved the lemma. \hspace{7cm} \qed

Moreover, the $(p,i)$-submodel is expressed by using $F_{n,\, \mu}$s, namely, 
\begin{lem}\label{lem:pi-sub}
the $(p,i)$-submodel is equivalent to 
\begin{equation}
 \sum_{\mu}{}':F_{p-i,\, \mu}\bar{F}_{i,\, \mu}:=0. 
\end{equation}
\end{lem}
\textit{proof}: \ We note that 
\begin{eqnarray*}
  && \sum_{\mu}{}'\ \partial_{\mu}^{p-i}(u^k)\partial_{\mu}^i(\bar{u}^l) \\
  &=& \sum_{\mu}{}'\ \sum_{j_1=1}^{p-i} B_{p-i,j_1}(g_1,\cdots,g_{p-i-j_1+1})
       \left(
        \frac{\partial}{\partial u}
       \right)^{j_1} (u^k) \\
  && \qquad \times 
      \sum_{j_2=0}^i B_{i,j_2}(\bar{g}_1,\cdots,\bar{g}_{i-j_2+1})
       \left(
        \frac{\partial}{\partial \bar{u}}
       \right)^{j_2} (\bar{u}^l) \\
  &=& \sum_{j_1=1}^{p-i} \sum_{j_2=0}^i j_1! j_2!
       \left(
       \begin{array}{cc}
       	 k \\
       	 j_1
       \end{array}
       \right)
       \left(
       \begin{array}{cc}
       	 l \\
       	 j_2
       \end{array}
       \right) \\
   && \qquad \times
       \sum_{\mu}{}'\ B_{p-i,j_1}(g_1,\cdots,g_{p-i-j_1+1}) 
         B_{i,j_2}(\bar{g}_1,\cdots,\bar{g}_{i-j_2+1}) 
          u^{k-j_1} \bar{u}^{l-j_2} \\
  && \quad \mbox{for} \quad k=1,\cdots, p-i, \ l=0,\cdots,i. 
\end{eqnarray*}
Because of this, the $(p,i)$-submodel holds if and only if
\begin{eqnarray*}
&& \sum_{\mu}{}'\ B_{p-i,j_1}(g_1,\cdots,g_{p-i-j_1+1}) 
    B_{i,j_2}(\bar{g}_1,\cdots,\bar{g}_{i-j_2+1})=0 \\
  && \quad \mbox{for} \quad j_1=1,\cdots, p-i, \ j_2=0,\cdots,i, 
\end{eqnarray*}
namely
$$
 \hspace{45mm}
 \sum_{\mu}{}':F_{p-i,\, \mu}\bar{F}_{i,\, \mu}:=0. 
 \hspace{45mm} \qed
$$
In view of (\ref{eqn:iden}) and the lemmas above, 
we can search an infinite number of conserved currents as follows; \\
For fixed $(p,i)$, we consider $\xi^{p-i}\bar{\xi}^i$ and 
$\xi^i\bar{\xi}^{p-i}$ in $\mathbf{C}[\xi, \bar{\xi}]$ corresponding to 
the $(p,i)$-submodel. 
Find the polynomials $p(\xi,\bar{\xi})$ such that
\begin{equation}
 (\xi+\bar{\xi})p(\xi,\bar{\xi})=
 \alpha \xi^i\bar{\xi}^{p-i}+\beta \xi^{p-i}\bar{\xi}^i 
  \qquad \mbox{for some} \ \alpha, \beta \in {\mathbf{C}} .
\end{equation}
Then we can decide it uniquely (up to constant). That is
\begin{equation}
 p(\xi,\bar{\xi})=\sum_{k=0}^{p-1-2i}(-1)^k \xi^{p-1-i-k}\bar{\xi}^{i+k}.
\end{equation}
Therefore, if we define the operator 
\begin{equation}
V_{(p,i),\, \mu} \equiv 
  \sum_{k=0}^{p-1-2i}(-1)^k :F_{p-1-i-k,\, \mu}\bar{F}_{{i+k},\, \mu}:,
\end{equation}
we obtain the next theorem.
\begin{thm}For $p=2,3,\cdots $ \ and \ $i=0,1,\cdots,[(p-1)/2]$,
\begin{equation}
 V_{(p,i),\, \mu}(f)  
 \label{eqn:5-1}
\end{equation}
are conserved currents for the $(p,i)$-submodel, 
where $f=f(u,\bar{u})$ is any function in $C^p$-class. 
\end{thm}
For example, corresponding to (\ref{eqn:2-0sub}), (\ref{eqn:3-0sub}), and 
(\ref{eqn:3-1sub}),
\begin{eqnarray}
 V_{(2,0),\, \mu}(f) &=& F_{1,\, \mu}(f)-\bar{F}_{1,\, \mu}(f) \nonumber \\
   &=& \partial_{\mu}u \frac{\partial f}{\partial u}
      -\partial_{\mu}\bar{u} \frac{\partial f}{\partial \bar{u}},
      \hspace{75mm}
\end{eqnarray}
\begin{eqnarray}
 V_{(3,0),\, \mu}(f) &=& F_{2,\, \mu}(f)
                    -:F_{1,\, \mu}\bar{F}_{1,\, \mu}:(f)
                    +\bar{F}_{2,\, \mu}(f) \nonumber \\
   &=& \partial_{\mu}^2 u \frac{\partial f}{\partial u}
      +(\partial_{\mu}u)^2 \frac{\partial^2 f}{\partial u^2}
      -\partial_{\mu}u \partial_{\mu}\bar{u} 
        \frac{\partial^2 f}{\partial u \partial \bar{u}}
      +\partial_{\mu}^2 \bar{u} \frac{\partial f}{\partial \bar{u}}
      +(\partial_{\mu}\bar{u})^2 \frac{\partial^2 f}{\partial \bar{u}^2},
      \nonumber \\
   && 
\end{eqnarray}
\begin{eqnarray}
 V_{(3,1),\, \mu}(f) &=& :F_{1,\, \mu}\bar{F}_{1,\, \mu}:(f) \nonumber \\
   &=& \partial_{\mu}u \partial_{\mu}\bar{u} 
        \frac{\partial^2 f}{\partial u \partial \bar{u}}.
      \hspace{83mm}
\end{eqnarray}
(\ref{eqn:5-1}) is a generalization of the conserved currents 
for the $p$-submodel in \cite{FHS2}.

 \section{Exact Solutions}
In spite of higher order equations, we find that the $(p,i)$-submodel 
has a B\"acklund-like transformation of solutions 
\begin{equation}
 v=f(u)=\sum_{i=0}^{\infty}f_i u^i 
  \qquad f:{\mathbf{C}} \longrightarrow {\mathbf{C}} :\mbox{\ holomorphic}
\end{equation}
with an infinite number of parameters $f_i \ (i=0,1,\cdots )$. 
This property is similar to the Grassmann submodel \cite{FHS1}.
\begin{thm}\label{thm:backlund}
If $u$ is a solution of the $(p,i)$-submodel, 
then, for any holomorphic function 
$f$, $v=f(u)$ is also a solution of 
the $(p,i)$-submodel. 
\end{thm}
\textit{proof}: \ 
Suppose that $u$ is a solution of the $(p,i)$-submodel. 
By lemma \ref{lem:pi-sub}, 
the $(p,i)$-submodel is equivalent to 
\begin{equation}
  \sum_{\mu}{}':F_{p-i,\, \mu}\bar{F}_{i,\, \mu}:=0, 
\end{equation}
namely
\begin{equation}
 \sum_{\mu}{}' B_{p-i,j}[u]B_{ik}[\bar{u}]=0
 \quad \mbox{for} \quad j=1,\cdots, p-i, \ k=0,\cdots,i. 
\end{equation}
Therefore, by using of (\ref{eqn:Bmat})
\begin{eqnarray*}
 \sum_{\mu}{}' B_{p-i,j}[f(u)]B_{ik}[\overline{f(u)}]
 &=&
 \sum_{\mu}{}' \sum_{n=j}^{p-i} \sum_{m=k}^i 
     B_{p-i,n}[u]B_{nj}[f] B_{im}[\bar{u}]B_{mk}[\bar{f}] \\
 &=& 0. \hspace{7cm} \qed
\end{eqnarray*}
For complex numbers $a_{\mu} \ (\mu =0,1,\cdots,n)$ with 
$\sum_{\mu}{}'a_{\mu}^{p-i}\bar{a}_{\mu}^i=0$, 
\begin{equation}
 u=a_0 x_0+\sum_{i=1}^n a_i x_i
\end{equation}
are clearly solutions of (\ref{eqn:4-4}). 
Therefore, we obtain the following corollary. 
\begin{cor}
Let $f$ be any holomorphic 
function. Then
\begin{equation}
 f(a_0 x_0+\sum_{i=1}^n a_i x_i)
\end{equation}
under
\begin{equation}
 \sum_{\mu}{}'a_{\mu}^{p-i}\bar{a}_{\mu}^i=0
\end{equation}
are solutions of the $(p,i)$-submodel.
\end{cor}

 \section{Discussion}
In this paper, we defined a new system of PDE
and constructed a B\"acklund-like transformation of solutions and 
an infinite number of conserved currents by using the Bell polynomials. 
In particular, we constructed the necessary differential operators 
by a natural correspondence of Bell polynomials with usual monomials. 

Our result is a generalization of that of \cite{FHS2} and gives 
a simpler method for constructing an infinite number of conserved currents. 
We remark that the extended Smirnov-Sobolev construction 
in terms of \cite{FHS2}, which is a method for constructing exact solutions, 
is also valid for our new system of PDE. 

It is important to know the relationship between the conserved currents 
and the exact solutions of our submodels. Therefore, we need to investigate 
symmetries, Poisson structures on our system of PDE. A study of these 
structures is in progress. 
 \section*{Acknowledgements}
The author is very grateful to Kazuyuki Fujii for helpful comments 
on an earlier draft on this paper and to Yasushi Homma for 
helpful suggestion in Theorem \ref{thm:backlund}. 
\bibliographystyle{plain}

\end{document}